\documentclass[sigconf]{acmart}

\usepackage[english]{babel}
\usepackage{blindtext}

\usepackage{tikz}
\usepackage{amsmath}
\usepackage{filecontents}
\usepackage[english]{babel}
\usepackage{graphicx}
\usepackage{subfigure}
\usepackage{soul,color}
\usepackage{lipsum}  
\usepackage{xcolor}
\usepackage{marginnote}
\usepackage{booktabs}
\usepackage{multirow}
\usepackage{pifont}
\usepackage{tabularx}
\usepackage{hyperref}
\usepackage{lipsum}
\usepackage{algorithm}    
\usepackage{algorithmic}  

\renewcommand\footnotetextcopyrightpermission[1]{} 
\setcopyright{none}

\acmDOI{}

\acmISBN{}

\acmConference[Submitted for review]{}
\acmYear{2025}

\acmPrice{}

\begin{document}
\title{Saflo: eBPF-Based MPTCP Scheduler for Mitigating\\Traffic Analysis Attacks in Cellular Networks}

\author{Sangwoo Lee, Liuyi Jin, Radu Stoleru}
\orcid{1234-5678-9012}
\affiliation{%
  \institution{Computer Science and Engineering, Texas A\&M University}
}


\begin{abstract}
This paper presents the $\underline{\textbf{saf}}$e sub$\underline{\textbf{flo}}$w (Saflo) eBPF-based multipath TCP (MPTCP) scheduler, designed to mitigate traffic analysis attacks in cellular networks. Traffic analysis attacks, which exploit vulnerabilities in Downlink Control Information (DCI) messages, remain a significant security threat in LTE/5G networks. To counter such threats, the Saflo scheduler employs multipath communication combined with additional security-related tasks. Specifically, it utilizes eBPF tools to operate in both kernel and user spaces. In the kernel space, the eBPF scheduler performs multipath scheduling while excluding paths disabled by the user-space programs. The user-space programs conduct security-related computations and machine learning-based attack detection, determining whether each path should be enabled or disabled. This approach offloads computationally intensive tasks to user-space programs, enabling timely multipath scheduling in kernel space. The Saflo scheduler was evaluated in a private LTE/5G testbed. The results demonstrated that it significantly reduces the accuracy of video identification and user identification attacks in cellular networks while maintaining reasonable network performance for users.
\end{abstract}


\maketitle

\section{Introduction}
Monitoring user activities through Downlink Control Information (DCI) poses a significant security risk in LTE and 5G networks. DCI messages contain critical resource allocation details, such as the timing and volume of data transmission between the base station and the user. While the Packet Data Convergence Protocol (PDCP) layer is responsible for providing encryption and integrity protection, it does not secure DCI messages, as DCIs are handled by the lower layers (i.e., PHY and MAC). Consequently, the base station transmits DCI in plain text, leaving it vulnerable to interception.

Several studies have investigated attack models that exploit this vulnerability. In \cite{aLTEr}, the authors developed a website fingerprinting attack using pcap traces from the LTE base station (i.e., evolved NodeB or eNB) to demonstrate the feasibility of DCI-based traffic analysis attacks. Later, in \cite{VideoID}, the authors implemented a video identification attack in real-world LTE networks. In this study, they intercepted DCI information from mobile network operators' (MNOs') eNBs and analyzed it using a convolutional neural network (CNN) to identify videos watched by users.

In \cite{5GSniffer}, a user identification attack was developed for real-world 5G networks. In this attack, the attacker periodically sends messages to a victim. Then, it monitors the resulting traffic patterns through DCI messages broadcasted by the 5G base station (i.e., Next Generation NodeB or gNB) to detect the victim's presence in the gNB coverage area. Although these studies have shown the feasibility of DCI-based traffic analysis attacks in LTE/5G networks, no research has yet addressed methods to mitigate them, to the best of our knowledge.

Meanwhile, traffic analysis poses a serious security threat to Tor networks, as the middle Onion routing node can serve as a vantage point \cite{Tor_tuto, deepFingerprinting, Click,Relation-CNN, Peek-a-Boo}. Accordingly, extensive research has focused on developing countermeasures to preserve anonymity in Tor networks. In \cite{Buflo, CS-Buflo, TAMARAW, WTF-PAD, walkie, Front}, the authors proposed countermeasures that introduce artificial delays or add padding data to confuse attackers. While these countermeasures effectively protect Tor networks from traffic analysis attacks, they introduce overhead due to the added delay and padding, making them unsuitable for LTE/5G networks, where quality of service (QoS) is essential.


Cadena \textit{et al.} \cite{TrafficSliver} and Henri \textit{et al.} \cite{HyWf} proposed a multipath-based countermeasure that reduces attacker accuracy by randomly splitting traffic across multiple paths. This approach incurs less overhead compared to delay- or padding-based methods, making it a promising solution. As multipath communication is already an integral part of cellular networks, referred to as \textit{“Access Traffic Steering, Switching, and Splitting (ATSSS)''} \cite{ATSSS, MPTCPin5G,ATSSS1,ATSSS2,MILCOM-atts}, the multipath-based approach also holds the potential for mitigating DCI-based traffic analysis attacks in cellular networks.

However, the countermeasures mentioned above were originally designed for Tor networks and are not applicable to cellular networks. Additionally, in Tor networks, multipath communication typically occurs over multiple routes on the same network interface, resulting in lower path heterogeneity. In contrast, LTE/5G networks commonly exhibit significant path heterogeneity.

In LTE/5G networks, multipath communication utilizes diverse network interfaces, such as LTE, 5G, WiFi, and potentially satellite communication in future 6G networks \cite{non-3gppTutorial}. This diversity makes path heterogeneity a key factor in multipath communication \cite{BLEST,ECF,xlink,Peekaboo,CellFusion,Converge}. Notably, path heterogeneity can lead to head-of-line (HoL) blocking \cite{MPTCP_scheduling}, which occurs when packets arrive out of order due to variations in path characteristics. This out-of-order delivery results in packet drops and significant delays in network queues. 

Although many multipath schedulers have been developed to prevent HoL blocking, they were not designed with traffic analysis attacks in mind. For example, BLEST \cite{BLEST} is considered one of the state-of-the-art schedulers for mitigating HoL blocking and has been adopted as the default multipath TCP (MPTCP) scheduler in the Linux kernel. However, we found that BLEST is insufficient to mitigate DCI-based traffic analysis attacks, allowing an attacker to achieve an accuracy of over 70\% in the video identification attack and 90\% in the user identification attack. Thus, there is a need to develop an MPTCP scheduler that mitigates DCI-based traffic analysis attacks while ensuring reasonable network performance for cellular network users.


In this paper, we present the $\underline{\textbf{saf}}$e sub$\underline{\textbf{flo}}$w (Saflo) scheduler, which is an eBPF(extended Berkeley Packet Filter)-based MPTCP scheduler to mitigate DCI-based traffic analysis attacks while maintaining the reasonable network performance. Specifically, the Saflo scheduler aims to defend against video identification attacks \cite{VideoID} and user identification attacks \cite{5GSniffer}, which have demonstrated real-world feasibility in LTE/5G networks.

The Saflo scheduler utilizes eBPF tools to operate across both kernel and user spaces. Using eBPF, the Saflo scheduler offloads security-related computations and machine learning-based attack detection to user-space programs. The Saflo scheduler consists of three main components: the eBPF scheduler in kernel space, and the subflow manager and attack detector in user space.

\textbf{The eBPF scheduler} employs BLEST-like scheduling to minimize HoL blocking while excluding subflows disabled by the subflow manager. Note that an individual path within an MPTCP connection is referred to as a subflow. \textbf{The subflow manager} periodically determines whether each subflow should remain enabled or be disabled. This decision is based on kernel parameters provided by the eBPF scheduler and detection results from the attack detector. To enhance security, the subflow manager introduces randomness into its decision-making process, generating unpredictable traffic patterns across the subflows. \textbf{The attack detector} periodically retrieves traffic information monitored by the eBPF scheduler and performs attack detection using CNN classifiers. These three components interact via reading and updating eBPF maps, shared memory, and files.

We evaluated the Saflo scheduler in terms of both security and network performance. The results show that it significantly reduces video identification attack accuracy from 96.9\% to 12.4\% and user identification attack accuracy from 99.4\% to 59.5\%. Additionally, it outperforms single-path cellular networks and the state-of-the-art approach in the Tor network \cite{TrafficSliver} in throughput, out-of-order packets, and completion time.

Our contributions are summarized as follows:
\begin{itemize}
    \setlength{\itemsep}{1pt} 
    \setlength{\parskip}{0pt} 
    \setlength{\parsep}{0pt}  
    \item We reproduce the DCI-based traffic analysis attacks and show that the existing MPTCP is not enough to defend against the attacks.
    \item We present the first study on a countermeasure against DCI-based traffic analysis attacks using multipath. To enhance network security, our countermeasure integrates complex operations (e.g., machine learning models) in user space into kernel-level operations using eBPF.
    \item We demonstrate that our countermeasure can mitigate the DCI-based traffic analysis attacks dramatically while it preserves reasonable network performance.
    \item We open-source the used scripts, datasets, and our countermeasure to encourage further research. \cite{Saflo-git}\footnote{The project will be open-sourced after publication.}.
\end{itemize}
This work does not raise any ethical issues.


\section{Background}
\begin{figure}
    \centering
    \includegraphics[width=0.65\linewidth]{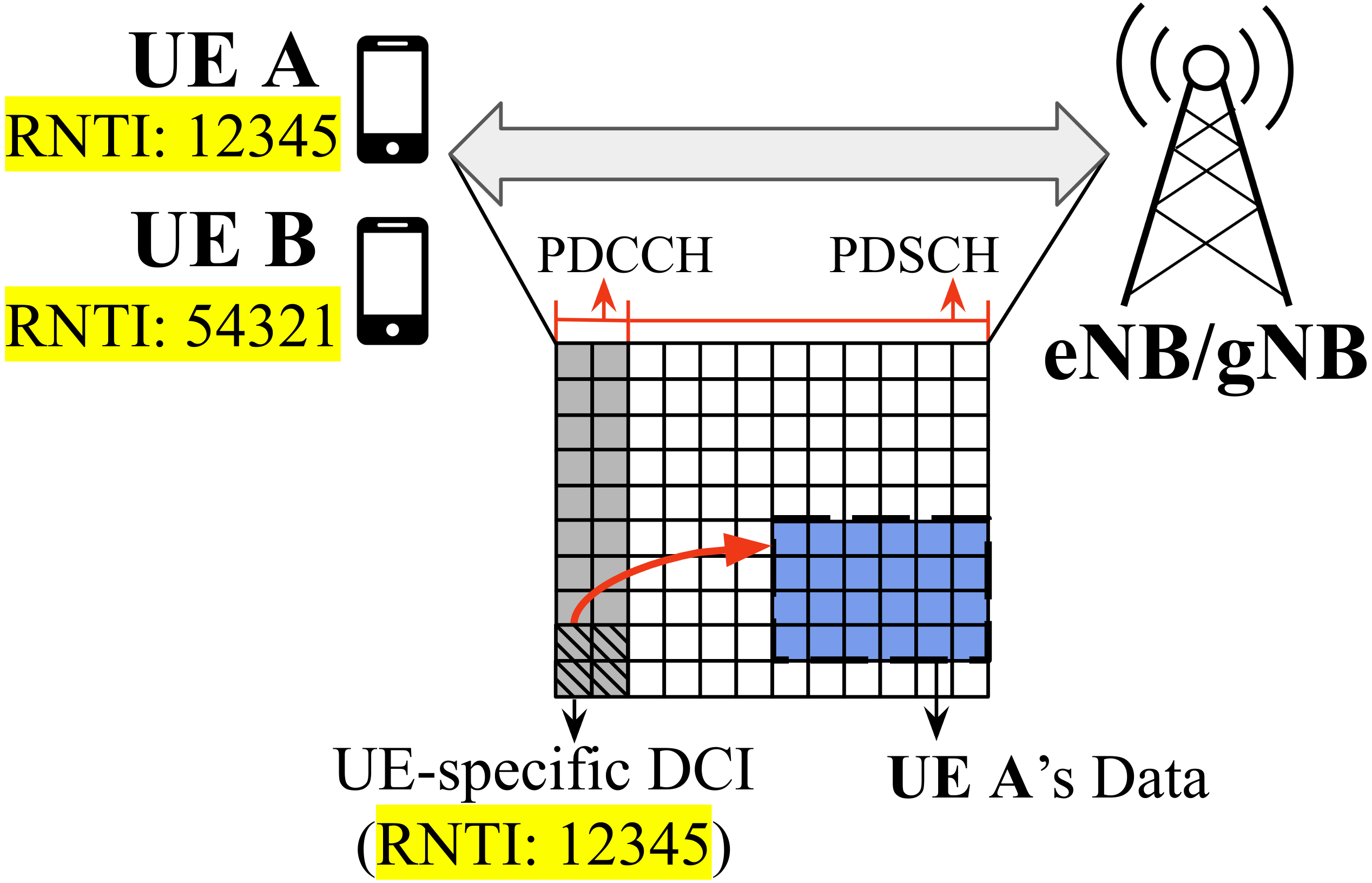}
    \caption{Downlink resource allocation Using DCI}
    \label{fig:enter-label}
\end{figure}
\subsection{Resource Allocation in LTE/5G}
Resource allocation in LTE/5G networks mainly involves the Physical Downlink Control Channel (PDCCH) and Downlink Control Information (DCI) \cite{pdcch}. Figure 1 illustrates downlink resource allocation using DCI. In the figure, the base station (eNB or gNB) assigns distinct Radio Network Temporary Identifiers (RNTIs) to two users, \textit{UE A} and \textit{UE B}. These RNTIs serve as temporary identifiers for communication and are used for addressing, signaling, and resource allocation.

Using the assigned RNTI, the base station transmits user-specific DCI via the PDCCH to notify \textit{UE A} about an upcoming downlink transmission. The PDCCH resides in the control region of each subframe in LTE or in the first few symbols of a slot in 5G NR. UEs decode the PDCCH within predefined search spaces, which can be either common (shared by all UEs) or UE-specific.

Once \textit{UE A} successfully decodes the DCI on the PDCCH, it retrieves resource allocation details, such as the allocation of physical resource blocks (PRBs), modulation and coding schemes (MCS), transport block sizes (TBS), and hybrid automatic repeat request (HARQ) parameters. Using this information, \textit{UE A} identifies the upcoming downlink transmission on the Physical Downlink Shared Channel (PDSCH) and receives the data.

Uplink resource allocation in LTE and 5G networks follows a similar process. The base station sends an uplink grant via the DCI on the PDCCH, specifying the PRBs and transmission parameters for the UE to use on the Physical Uplink Shared Channel (PUSCH). With this information, the UE transmits its uplink user data on the PUSCH.


\subsection{Threats of DCI-based Traffic Analysis}
\textbf{Threats Overview.}
In LTE/5G networks, the PDCP layer is responsible for encryption and integrity protection. However, DCI messages are managed by the PHY and MAC layers, which operate below the PDCP layer. Consequently, DCI messages are transmitted in plain text, making them vulnerable to traffic analysis attacks. 

In general, DCI-based traffic analysis attacks aim to identify whether a user is engaged in a specific targeted activity \cite{aLTEr,VideoID,5GSniffer}. These attacks typically involve three main steps. 

In the first step, the attackers collect traffic metadata from DCI messages generated during the targeted activity. To do this, the attackers connect to the target network, perform the specific activity, and intercept the associated DCIs. Then, relevant metadata is extracted from the captured DCIs and stored for later analysis.

In the second step, the attacker trains a classifier using the collected metadata. This classifier determines whether certain traffic metadata corresponds to the targeted activity. The classifier can take various forms, such as a neural network, support vector machine (SVM), or k-nearest neighbors (k-NN) model with dynamic time warping (DTW).

In the last step, the attacker monitors the target network, intercepts new DCI messages, and uses the trained classifier to identify whether a user is performing the targeted activity.

\textbf{Key Challenges and Solutions.} 
When implementing DCI-based traffic analysis attacks, the attacker may encounter several technical challenges. First, the attacker should handle the user's changing RNTIs. When a UE stops using the network, it switches to the Radio Resource Control (RRC) idle state, releasing its current RNTI. Upon resuming network activity, the UE transitions back to the RRC connected state, and the eNB/gNB assigns it a new RNTI. The RRC inactive time period can range from 100 ms to several seconds. Thus, the attacker must either find a way to track the user's changing RNTIs or force the user to retain the same RNTI.

If the user's public ID (e.g., messenger ID or phone number) is known, the attacker can force the user to retain the same RNTI during the attack. In \cite{5GSniffer}, the attacker achieved this by periodically sending stealthy messages at 2.5-second intervals. These malicious messages were not visible to the user due to a vulnerability in messenger apps. 

Furthermore, identity mapping attack can also be exploited to track user RNTIs, as demonstrated in \cite{aLTEr, VideoID}. When a UE transitions from the RRC idle state to the RRC connected state, it initiates the process by sending a \textit{Random Access Channel (RACH) request} to the eNB or gNB. In response, the eNB/gNB sends a \textit{RACH response}, which includes a newly assigned RNTI. After that, the UE and eNB/gNB exchange the \textit{RRC setup request} and \textit{RRC setup response}, where both the UE’s Temporary Mobile Subscriber Identity (TMSI) and the assigned RNTI are contained.

The TMSI, assigned by the core network, typically has a longer validity period (e.g., several hours or days) compared to the shorter-lived RNTI\cite{overshadow, AdaptOver}. Notably, the \textit{RACH request/response} and \textit{RRC setup request} are transmitted in plain text since the eNB/gNB has not yet identified the UE. Similarly, the \textit{RRC setup response} is often in plain text but may be encrypted depending on the network configuration. By intercepting these unencrypted messages, an attacker can link the TMSI to the RNTI, allowing them to track the user's RNTIs over time. This attack was initially developed for LTE networks but remains effective in 5G networks.

Carrier Aggregation (CA) introduces another challenge for attackers. When CA is enabled, user data is transmitted through both a primary serving cell (PCell) and a secondary serving cell (SCell), making it difficult for the attacker to accurately capture traffic metadata. Since the PCell and SCell share the same RNTI for a UE, a straightforward solution is to deploy multiple sniffers to monitor all eNBs/gNBs in the cell group. However, Bae \textit{et al.} \cite{VideoID} proposed an alternative approach using only a single sniffer. Their method leverages the fact that the PDCP packet headers, which are not encrypted, include data sequence numbers that are shared between the PCell and SCell. By decoding data sequence numbers on the PCell's PDSCH, the attacker can infer the amount of traffic transmitted through the SCell.

\subsection{Multi-access LTE/5G Networks}
In this research, we aim to develop a countermeasure against DCI-based traffic analysis attacks by leveraging multi-path communication that is already an integral part of LTE/5G network systems \cite{3gpp_TS_23.501, ATSSS, 5G-MANTRA}.

Figure 2 shows the system architecture of multi-access LTE/5G networks. In this setup, UE can access MNOs' services through both traditional 3GPP access networks (i.e., eNB/gNB) and non-3GPP access networks (i.e., WiFi). The larger blue boxes represent the 5G core network functions, while the smaller yellow boxes represent the LTE core network functions. Both 5G and LTE functions grouped together perform similar tasks at a high level. For simplicity, we focus on explaining the 5G core network functions.

\begin{figure}
    \centering
    \includegraphics[width=0.95\linewidth]{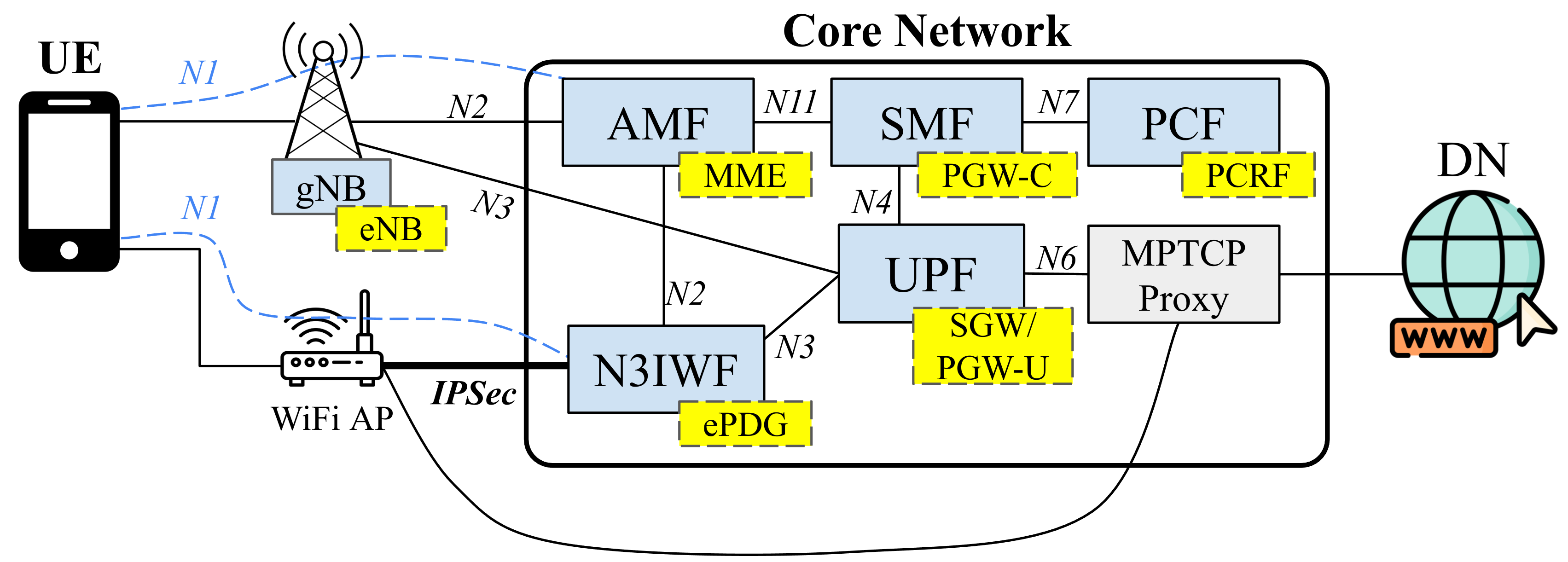}
    \caption{Multi-access LTE/5G networks}
    \label{fig:enter-label}
\end{figure}

In 5G networks, essential network functions (NFs) include the Access and Mobility Management Function (AMF), which handles user device authentication and mobility management; the Policy Control Function (PCF), responsible for implementing security, quality of service, and charging policies; the Session Management Function (SMF), which oversees user sessions and assigns IP addresses; and the User Plane Function (UPF), which manages data traffic by performing packet routing, forwarding, and traffic handling.

The Non-3GPP Interworking Function (N3IWF) is an optional feature that allows a UE to connect to the 5G core network via non-3GPP access networks (e.g., WiFi). After the UE's authentication, the N3IWF and UE can derive a security key for non-3GPP access ($K_{N3IWF}$) from the root key \cite{3gpp_TS_33.501}. Using $K_{N3IWF}$, the N3IWF and UE establish an IPSec tunnel to encrypt traffic over the non-3GPP access network. After establishing the IPSec tunnel, the N3IWF relays non-3GPP traffic between UE and UPF.

The MPTCP proxy allows UEs to utilize MPTCP across different wireless network interfaces for any available services on the internet. In 3GPP Release 16, MPTCP was chosen as the primary transport layer protocol for multi-path communication, also referred to as \textit{“ATSSS''} operations \cite{ATSSS, MPTCPin5G}. However, real-world adoption of MPTCP has been limited, restricting its use for general-purpose internet services.

A promising solution to address this limitation is deploying an MPTCP proxy in the core network \cite{MPTCP_proxy1,MPTCP_proxy2, 5G-MANTRA}. The proxy facilitates seamless multi-path communication by splitting downlink traffic across multiple paths and reassembling uplink traffic toward external services. Furthermore, the MPTCP proxy handles multi-path scheduling for downlink traffic, making it essential to implement effective scheduling mechanisms within the proxy. 

When the UE connects to the MPTCP proxy, it has two options: connection through the N3IWF or direct connection. From a security perspective, the main difference lies in the authentication process. A direct connection may rely on a separate key exchange, such as TLS or SSH. In contrast, a connection through the N3IWF can provide better security using the IPSec tunnel established with  $K_{N3IWF}$, derived from the root key shared between the core network and the SIM card.


\section{Attack Reproduction}
\begin{figure}
    \centering
    \includegraphics[width=1\linewidth]{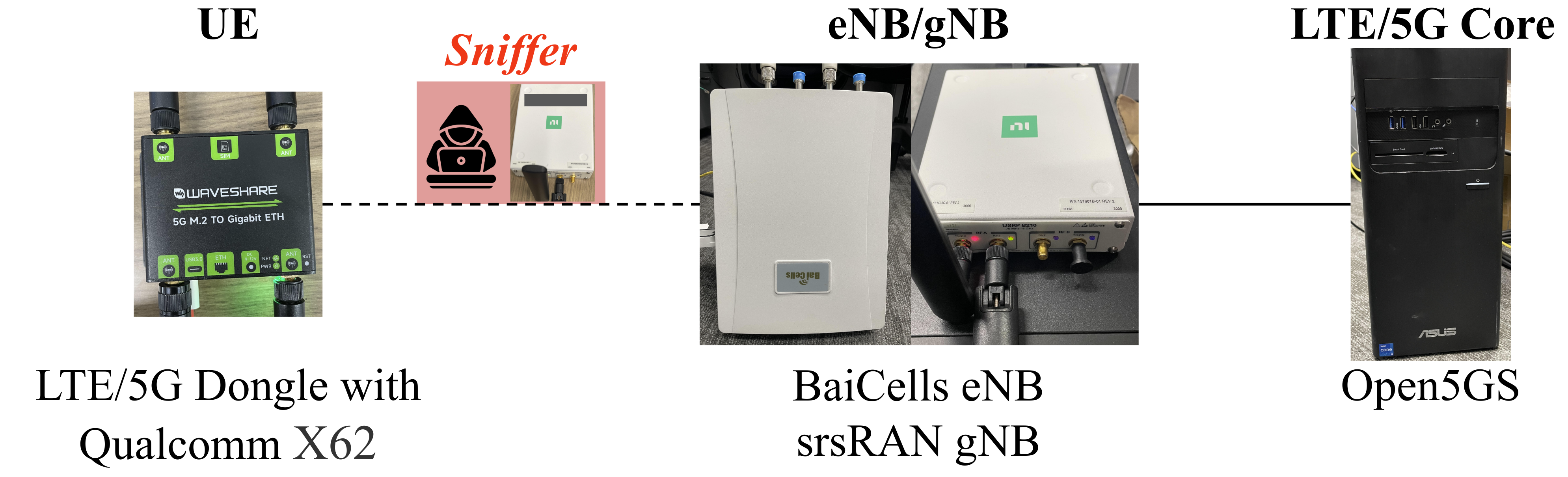}
    \caption{Private LTE/5G testbed}
\end{figure}

For the development and evaluation of the countermeasure, we first replicated two DCI-based traffic analysis attacks (i.e., video identification and user identification) in our LTE/5G testbed. Figure 3 illustrates our testbed. In this setup, we deployed the open-source LTE/5G core network, open5GS \cite{open5gs}, on a PC equipped with an Intel i5-12400 CPU. The LTE eNB was a commercial Baicells eNB, while the 5G gNB utilized the open-source srsRAN-5G software \cite{srsRANdoc}, running on the same PC as the core network, along with a USRP B210 software-defined radio (SDR). The eNB and gNB operated in FDD mode on Band 5 and Band N3, respectively. The UE was an LTE/5G dongle with a Qualcomm Snapdragon X62, connected to a Linux laptop. Additionally, we deployed open-source DCI sniffers, LTESniffer \cite{LTEsniffer} for LTE networks and 5GSniffer \cite{5GSniffer} for 5G networks, both using another USRP B210. 

It is important to note that we were unable to deploy the attacks against real-world LTE/5G networks in our area. Current open-source sniffers lack support for several key features, including TDD, Multiple-Input Multiple-Output (MIMO), PDSCH sniffing, and wider bandwidth capabilities. The MNOs' eNBs and gNBs in our area operate using either TDD or MIMO, and the current open-source sniffers were unable to sniff DCI from them stably. These limitations, along with our experiences, are detailed in Section 6.

\subsection{\textbf{Video Identification Attack}}
We reproduced the attack described in \cite{VideoID}, which aims to identify the video being watched by users. This attack leverages the unique on-off traffic patterns for each video generated by HTTP Adaptive Streaming (HAS) \cite{VideoID, Beauty, EncryptedHTTP}. The top of Figure 4 illustrates an example of these on-off traffic patterns during YouTube streaming. It shows that data is transmitted at specific intervals (on periods), while no traffic occurs between these transmissions (off periods).

\begin{figure}
    \centering
    \includegraphics[width=0.75\linewidth]{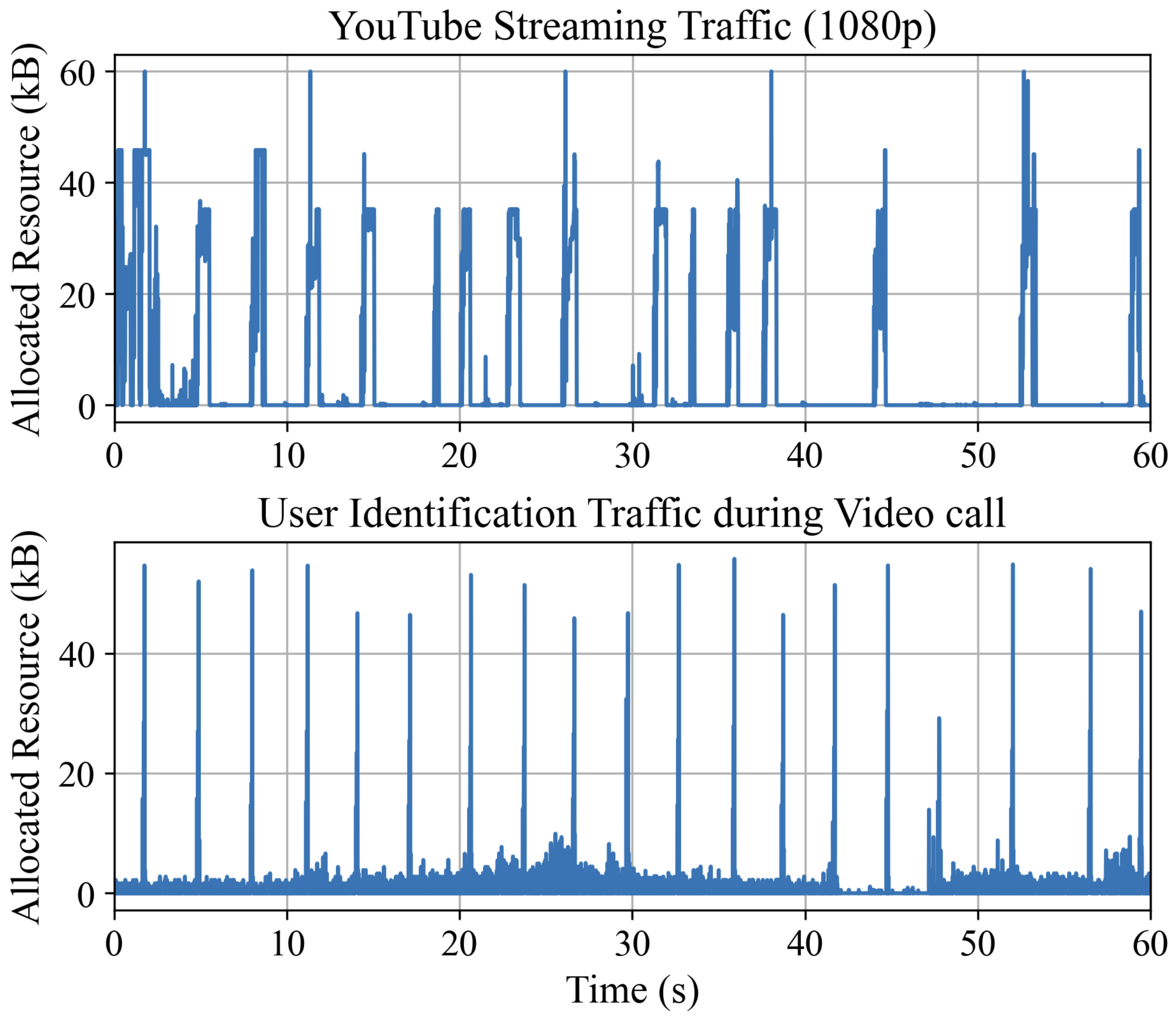}
    \caption[Example of Traffic Patterns]{Example of traffic patterns during attacks captured using LTESniffer}
\end{figure}

For implementation, we first played the Top 50 YouTube music videos 20 times each over an LTE network at 1080p resolution \cite{Top-youtube-videos}. During streaming, we collected DCI information using a sniffer to create a traffic meta-dataset. Unlike \cite{VideoID}, which used a commercial LTE sniffer, we utilized the open-source LTESniffer. We then used 80\% of this dataset to train a CNN classifier developed with TensorFlow Keras, following the same model structure as in \cite{VideoID}. The classifier was designed to take a 120-second traffic trace as input and classify it into one of 50 video labels. The training was performed over 15 epochs using a learning rate of 0.001. The remaining 20\% of the dataset was set aside to evaluate the attack accuracy.

In addition, we extended the attack to 5G networks. Initially, we collected DCI information using the open-source 5GSniffer. However, 5GSniffer \cite{5GSniffer} faced significant limitations, particularly SDR overflow issues that affected its ability to decode DCIs. For example, during a one-minute YouTube video stream that involves approximately 13,000 DCIs, 5GSniffer interpreted between 1,000 and 10,000 DCIs. Consequently, the accuracy of the video identification attack using 5GSniffer was below 70\% when classifying five video labels.

To address these limitations, we assumed that the attacker could access DCI information from the gNB log, making the attacker significantly more powerful. This assumption ensures that if our countermeasure can thwart an attacker leveraging perfect DCI information, it will also be effective against attackers using sniffers. Thus, this approach is better suited to achieving our objective.

Based on this assumption, we collected DCI information directly from the logs of the srsRAN gNB while streaming Top 50 YouTube music videos 20 times each. Using these data, we constructed a traffic meta-dataset for the 5G network attack. The model structure and training process were identical to those used for the LTE network attack. 

As a result, the video identification attack achieved an accuracy of \textbf{96.93\%} in the LTE network, while it achieved \textbf{98.41\%} in the 5G network. The attack in 5G networks showed higher accuracy because it utilized perfect DCI information, whereas LTESniffer in LTE networks missed some DCIs due to over-the-air limitations. 
\subsection{\textbf{User Identification Attack}}
We implemented the user identification attack based on the approach outlined in \cite{VideoID}. The attack operates on a simple principle: the attacker periodically transmits data and detects the resulting data patterns using DCI information. The bottom part of Figure 4 illustrates an example of periodic traffic generated by the attacker while the user is on a video call. Although residual traffic from the video call is present, the attacker's traffic is short and bursty, making it easy to distinguish.

For implementation, we transmitted a message with a 1 MB image via the Discord app to the user every 2.5 seconds while recording the DCI information. Note that we used regular direct messages, whereas the authors of \cite{5GSniffer} used stealthy messages that are hidden from the target. However, the use of non-stealthy messages does not compromise the validity of our countermeasure evaluation as both methods achieve the same goal: generating a specific traffic pattern for the user to receive.

 During the DCI information collection, we considered the seven scenarios as follows: \textbf{(1)} the user only receives traffic from the attacker, \textbf{(2)}  the user watches a 1040p YouTube video, \textbf{(3)} the user watches a 1040p YouTube video while also receiving traffic from the attacker, \textbf{(4)} the user is on a voice call, \textbf{(5)} the user is on a voice call while receiving traffic from the attacker, \textbf{(6)} the user is on a video call, and \textbf{(7)} the user is on a video call while receiving traffic from the attacker.


We collected a 10-minute DCI trace for each scenario, resulting in a total of 70 minutes of trace data across all scenarios. From this data, 80\% was used to train the CNN binary classifier, while the remaining 20\% was reserved for evaluation. The classifier processes 10-second segments of DCI information to determine whether the traffic belongs to the target or not.

While we employed LTESniffer for attacks on LTE networks, we utilized gNB logs for 5G networks instead of relying on 5GSniffer. In \cite{5GSniffer}, the authors noted that \textit{``we stored the raw IQ samples until the successful conclusion of each experiment.''} We interpreted the term \textit{``successful''} to indicate that a sufficient number of DCIs were decoded. Using 5GSniffer, the maximum attack accuracy was 76.09\% without filtering, whereas the accuracy improved to 82.14\% by filtering out traces where the number of decoded DCIs fell in the lower 50\% for each scenario category. The accuracy after filtering matches the results reported in \cite{5GSniffer}.

On the other hand, when using the DCI information from gNB logs, the attack accuracy was stable and higher than that achieved with 5GSniffer, due to the availability of perfect DCI information. Since our objective is to develop effective countermeasures, a more powerful attack serves this purpose better. Therefore, we used the user identification attack using gNB logs as the baseline.

Consequently, the user identification attack achieved an accuracy of \textbf{98.53\%} in the LTE network, while it achieved \textbf{99.84\%} in the 5G network.

\section{Proposed Countermeasure}
This section details the operation of the Saflo scheduler, specifically designed to mitigate DCI-based traffic analysis attacks in LTE/5G networks. The Saflo scheduler is intended for deployment on machines running the MPTCP proxy, where it manages downlink traffic scheduling. It extends the BLEST scheduler's implementation within the Linux kernel, incorporating significant modifications achieved through eBPF.

The use of eBPF offers two key benefits. First, it facilitates interaction between user-space programs and kernel-space operations, allowing complex tasks such as generating randomized traffic patterns and performing machine learning-based attack detection to be offloaded to user-space programs. This design ensures efficient and timely execution of kernel operations. Second, eBPF simplifies the deployment of our scheduler. With a modern Linux kernel (e.g., Linux kernel 6.13), users can easily attach the scheduler using \textit{bpftool}, activate it with a single command (\textit{sysctl net.mptcp.scheduler=bpf\_saflo}), and run the corresponding user-space program.

\begin{figure*}[h]
    \centering
    \includegraphics[width=0.95\linewidth]{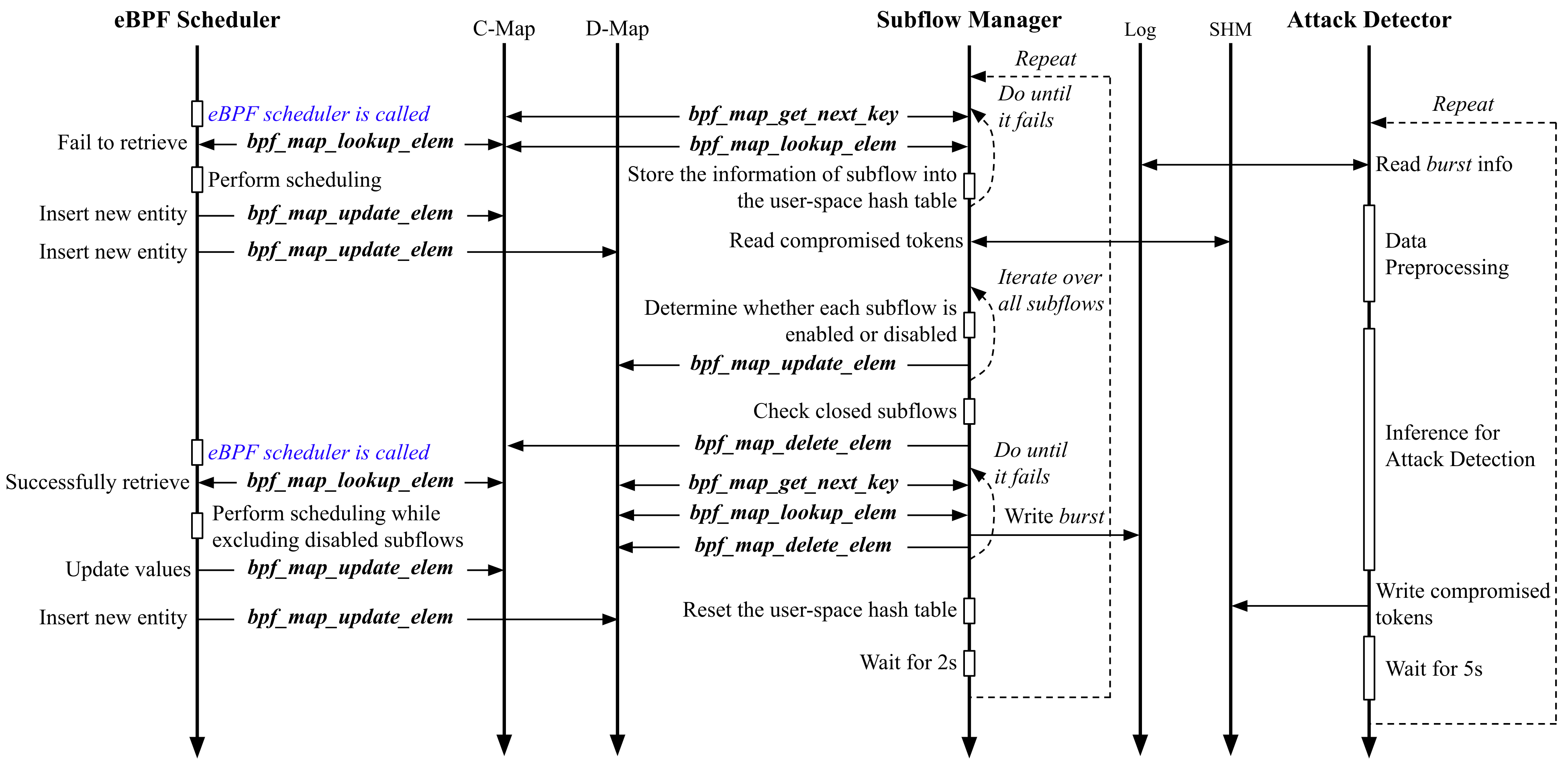}
    \caption{Overall operation of Saflo scheduler}
\end{figure*}

We first provide an overview of BLEST, which serves as the foundation for the Saflo scheduler. We then describe the design and implementation of the Saflo scheduler in detail.

\subsection{Foundation: BLEST Scheduler}
BLEST, the default MPTCP scheduler in the Linux kernel, is recognized for its effectiveness in HoL blocking. Its key operational principle lies in calculating the time required for scheduled data to be transmitted, rather than relying on simple network parameters like round-trip time (RTT). Specifically, BLEST estimates the time required to transmit data from the moment it is assigned to a subflow until the transmission is completed. This estimation considers each subflow's RTT, the size of the congestion window (CWND), and the amount of data remaining in the queue. Based on these factors, BLEST assigns data to the subflow with the shortest estimated transmission time. This strategy ensures that data arrives at the receiver in order, significantly reducing the likelihood of HoL blocking.

The implementation of BLEST in the Linux kernel is slightly different from the approach described in the original paper. Instead of calculating the expected data transmission time, the Linux kernel version computes the \textit{linger\_time} for each subflow, which is defined as $\frac{queued\_memory}{pacing\_rate}$, where \textit{queued\_memory} is the amount of data waiting in the transmission queue for a specific subflow, and \textit{pacing\_rate} is the rate at which data packets are sent over the network. After the calculation of \textit{linger\_time}, data is assigned to the subflow with the shortest \textit{linger\_time}. This simplified approach leverages kernel parameters to streamline scheduling decisions while still achieving BLEST's primary goal.



\subsection{Design of Saflo Scheduler}
Figure 5 depicts the overall operation of the Saflo scheduler, which comprises three components: the eBPF scheduler, the subflow manager, and the attack detector.

\textbf{eBPF Scheduler:} On the left top of Figure 5, there is a demand for scheduling MPTCP data, and the eBPF scheduler is called for the first time. Note that the eBPF scheduler functions as a callback, executed whenever the kernel needs to schedule MPTCP data. After the eBPF scheduler is called, it first attempts to retrieve the entity associated with the given MPTCP connection from the Control eBPF map (i.e., C-map). In the figure, the first attempting fails since no entity exists for the given MPTCP connection at the moment. Then, it performs scheduling by selecting the subflow with the shortest \textit{linger\_time}. 

Subsequently, the eBPF scheduler inserts new entities into the C-map with the information on subflows' \textit{linger\_time}, \textit{queued\_memory}, and \textit{pacing\_rate}. Lastly, it writes the new entity to the detection eBPF map (i.e., D-map), including information about the estimated traffic volume generated by this scheduling (i.e., \textit{burst} value in kernel code) and the current timestamp. Please note that the eBPF maps can be accessed by both the kernel side and user-space side through the \textit{libbpf} APIs such as \textit{\textbf{bpf\_map\_lookup\_elem}}, \textit{\textbf{bpf\_map\_update\_elem}}, and \textit{\textbf{bpf\_map\_delete\_elem}}.

When the eBPF scheduler is called for the second time, it successfully retrieves the entity associated with the given MPTCP connection from the C-map. It then performs scheduling in the same way while excluding any subflows marked as disabled in the retrieved entity. The entity includes a boolean variable for each subflow, indicating whether the subflow is enabled or disabled. Then, since the subflow information is updated during the scheduling process, the eBPF scheduler updates the values in the C-map entity with the new subflow information and writes the \textit{burst} value with timestamp to the D-map. 

\textbf{Subflow Manager:} The subflow manager repeats its operations at predefined intervals (i.e., 2 seconds as default). The operations of the subflow manager are performed in three steps.

First, it retrieves all entities from the C-map and stores them in a user-space hash table. The user-space hash table organizes subflow information by tokens, with each token representing an individual MPTCP connection. This structure enables the subflow manager to manage all subflows by token during subsequent operations. Then, it reads the shared memory (SHM) to check if any compromised MPTCP connections have been reported by the attack detector. 

In the next step, the subflow manager iterates through the subflows by each token to decide which subflows to enable or disable. If any compromised MPTCP connections are detected, it disables the subflows over LTE/5G networks to mitigate the user identification attack. It is important to note that we opted to send all traffic over WiFi instead of closing the compromised MPTCP connection. This decision was made because the messenger app (e.g., Discord) automatically recreates the MPTCP connection (socket) in less than three seconds, allowing the attacker to resume sending traffic. In addition, when both cellular and WiFi subflows are disabled, the kernel handles the traffic as plain TCP and transmits it to the user regardless.

For other regular subflows, it applies a weighted random decision, where subflows with shorter \textit{linger\_time} have a higher probability of being enabled. Once the subflow states are determined, the subflow manager updates the C-map by modifying a single boolean variable in each entity to indicate whether the corresponding subflow is enabled or disabled. After the update, it examines the current list of MPTCP connections and removes entities corresponding to closed MPTCP connections' tokens from the C-map.

As a final step, the subflow manager reads all entities in the D-map and write them into the detection log (i.e., Log in the figure) file. After writing, it deletes the entities from the D-map for further use. Additionally, the user-space hash table is reset to prepare for updates in the next interval. After completing these steps, the subflow manager waits for the predefined interval before repeating the operation.

\textbf{Attack Detector:} The attack detector periodically performs attack detection and reports the results to the subflow manager via the shared map. It begins by reading the accumulated traffic information from the detection log file and preprocessing it into time series data.

After preprocessing, the detector employs two CNN binary classifiers for attack detection. The primary classifier identifies the MPTCP connection (i.e., socket) responsible for delivering main malicious data, such as images or files. Meanwhile, the secondary classifier focuses on detecting the MPTCP connection that transmits text or signaling information. This design stems from an observed pattern: malicious messages utilize two MPTCP sockets. Attempts to detect both sockets with a single classifier resulted in reduced detection accuracy.

When an attack is detected, the detector writes tokens corresponding to the compromised MPTCP connections into the shared memory. This entire process is executed every 5 seconds.

\subsection{Implementation of Saflo Scheduler}
\subsubsection{eBPF Scheduler}
The eBPF scheduler is an eBPF program written in C using \textit{libbpf}, featuring custom functions for MPTCP scheduling. It is compiled with \textit{clang} and hooked into the kernel space. Modern Linux kernels provide a hooking point for the MPTCP scheduler via the \textbf{struct\_ops} feature. In other words, the functions used for MPTCP scheduling are organized within a predefined structure, \textbf{mptcp\_sched\_ops}, allowing the kernel to utilize any set of functions implemented in this format. Once the eBPF scheduler is integrated into the kernel, users can activate it by selecting the desired scheduler using the \textit{sysctl} command.

The eBPF scheduler extends the BLEST scheduler with two key modifications. First, it excludes subflows disabled by the subflow manager. Second, it updates the eBPF map entities (C-map and D-map) to facilitate subflow control and attack detection. A more detailed explanation of the scheduling operation is provided in the Appendix.

\subsubsection{Subflow Manager}
The subflow manager is implemented in C using \textit{libbpf}, similar to the eBPF scheduler, but it operates in user space. It regularly performs a set of primary operations at predefined time intervals. 

The first operation involves retrieving the existing MPTCP subflow information from the C-map and organizing it into a user-space hash table. It is important to note that while a single MPTCP connection can involve multiple subflows, each entry in the C-map represents a single subflow. Consequently, the subflow manager should retrieve all entries from the C-map and reorganize them by their corresponding MPTCP connections using the user-space hash table.

The key in the user-space hash table is the token that uniquely represents a single MPTCP connection. The corresponding value is a user-defined structure, \textbf{SFs}, which contains information about the subflows associated with the MPTCP connection. Specifically, this structure includes the total number of subflows, an array of the subflows' local IDs, an array of the subflows' remote IDs, and the sum of the subflows' \textit{linger\_time} values. The combination of a local ID and a remote ID uniquely identifies each subflow, while the other information is used later to determine whether the subflow is enabled.

\begin{algorithm}[t]
\definecolor{lightblue}{RGB}{173, 216, 230} 
\sethlcolor{lightblue}
\caption{Decision for subflow state}
\begin{algorithmic}[1]
\REQUIRE \textit{tocken}: a token of target MPTCP connection, \\\textbf{SFs}: a structure that includes information about subflows, $P_{min}$ and $P_{max}$: The minimum and maximum allowable probability values for enabling a subflow, respectively
\FOR{$i = 0$ \TO \textbf{SFs}.\textit{sf\_num}$-1$}
    \STATE \textbf{map\_entity} = Read map entity of which key is (\textit{token},\\ \;\;\;\;\;\;\;\;\;\;\;\;\;\;\;\;\;\;\;\;\;  \textbf{SFs}.\textit{localID}[i], \textbf{SFs}.\textit{remoteID}[i]) 
    \STATE /* Disable compromised cellular subflow*/
    \IF{\textbf{map\_entity}.\textit{safe} = False}
    \STATE \textbf{map\_entity}.\textit{enabled} = False
    \STATE Update C-map with \textbf{map\_entity}
    \STATE \textbf{Continue}
    \ENDIF
    \STATE /* Do weighted random Decision */
    \STATE $p = 1 - (\textbf{map\_entity}.\textit{linger\_time}/\textbf{SFs}.lt\_sum)$
    \STATE $p =$ min(max($p$, $P_{min}$), $P_{max}$)
    \STATE $th = \text{Generate a random float between 0 and 1.}$
    \IF{$p>th$} 
        \STATE \textbf{map\_entity}.\textit{enabled} = True
    \ELSE
        \STATE \textbf{map\_entity}.\textit{enabled} = False
    \ENDIF
    \STATE Update C-map with \textbf{map\_entity}
\ENDFOR
\end{algorithmic}
\end{algorithm}

After organizing the subflow information in the user-space hash table, the subflow manager checks whether the attack detector has identified the user identification attack. If an attack is detected, the subflow manager retrieves the compromised tokens from the shared map. It then updates the \textit{safe} value to False for the C-map entity associated with the compromised token and the remote ID of the cellular network.

It is important to note that the cellular network's subflow can be identified by its remote ID, as this represents each network interface on the receiver side. In the current version, the subflow manager determines the remote ID of the cellular network by cross-referencing the UE's IP address from the Open5GS AMF log with the IP address of the first subflow associated with remote ID 0, obtained from the output of \textit{ss --mptcp -i}. Alternatively, the user can manually specify the remote ID for the WiFi interface.


After marking the cellular subflow with compromised tokens, the subflow manager determines whether each subflow should be enabled, as detailed in Algorithm 1. The algorithm iterates through all subflows associated with a single MPTCP connection. During each iteration, it checks whether a subflow is safe by checking the \textit{safe} field in the map entity. If the \textit{safe} field is False, it disables the subflow.

If the subflow is safe, the subflow manager performs the weighted random decision to ensure that the faster subflow is assigned a higher probability of being enabled. It first calculates the weighted probability for the subflow, as shown in line 10. To prevent the probability from becoming excessively high or low, upper and lower bounds ($P_{max}$ and $P_{min}$) are applied at line 11. Then, a random float between 0 and 1 is generated as a threshold. If the weighted probability exceeds this threshold, the algorithm sets \textbf{map\_entity}.\textit{enabled} to True, enabling the corresponding subflow. Otherwise, it sets \textbf{map\_entity}.\textit{enabled} to False. Finally, the result of the weighted random decision is updated in the C-map (line 18).

After completing the weighted random decision for all subflows, the subflow manager checks the current MPTCP connections through \textit{ss} command and removes C-map entries for closed connections. Subsequently, it reads all entities in the D-map and write its information into the detection log file.
The key of D-map is a kernel time stamp and token, and the value is the kernel parameter, called \textit{burst}, which is the estimated volume of transmission during the scheduling process. The D-map entity written in the detection log is deleted from the D-map to make space for further use.



Once the operations are completed, the subflow manager cleans up the user-space hash table to prepare for the next time interval and then waits for the interval to elapse before repeating the operations.

\subsubsection{Attack Detector}

The attack detector is a Python script that analyzes a detection log file to identify attacks using CNN classifiers. It preprocesses the log into a 10-second time series of \textit{burst} values for each token. These traces are fed into primary and secondary CNN classifiers, which output a score between 0 (regular traffic) and 1 (attack traffic). The primary classifier identifies the socket transmitting malicious file data, while the secondary classifier detects the socket sending signal and text. The primary and secondary classifiers achieved detection accuracies of 98\% and 88\%, respectively. To minimize false detections, the secondary classifier's results are ignored when the primary classifier detects no attacks. These CNN classifiers are stored in lightweight TFLite format, with a combined size of 10.6 MB. The structure of CNN classifiers is detailed in the Appendix.

Since each token represents an MPTCP connection, the CNN classifier identifies the malicious sockets carrying the attacker’s traffic. This approach ensures that only the traffic from the malicious socket is routed through the WiFi subflow, while regular traffic continues to utilize the multipath. 

After the CNN classifier performs inference, the attack detector generates a list of tokens corresponding to the identified attacker’s traffic and writes it into the shared map. This process is repeated every 5 seconds.



\section{Evaluation}
We evaluated the Saflo scheduler in our private LTE/5G testbed, as detailed in Section 3, focusing on three key aspects: mitigating video identification attacks, mitigating user identification attacks, and network performance. We compared the Saflo scheduler with singlepath, the BLEST scheduler, and the random distribution (RD) scheduler. The RD scheduler is an eBPF MPTCP scheduler that replicates the state-of-the-art approach used in the Tor network \cite{TrafficSliver}.

\subsection{Mitigation of Video Identification}
\textbf{Comparison with existing MPTCP}: To evaluate the mitigation of the video identification attack, we first collected DCI information under an MPTCP connection using the Saflo and BLEST schedulers while streaming the top 50 YouTube music videos on both LTE and 5G networks. Please note that any application on Linux can enable MPTCP using \textit{mptcpize}. Each video was played 20 times for 2.5 minutes. Eventually, we generated six datasets, each containing 2,500 minutes of DCI traces, amounting to a total of 15,000 minutes. These datasets are summarized as follows:
\begin{itemize}
    \item \textbf{Saflo-LTE} and \textbf{Saflo-5G}: Collected under MPTCP with the Saflo scheduler, using two subflows over cellular and WiFi.
    \item \textbf{BLEST-LTE} and \textbf{BLEST-5G}: Collected under MPTCP with the BLEST scheduler, using two subflows over cellular and WiFi.
    \item \textbf{Single-LTE} and \textbf{Single-5G}: Collected under single-path connections, exclusively using the LTE network and 5G network, respectively. The attack accuracies reported in Section 3.1 are derived from these datasets.
\end{itemize}
The attack model is identical to the one described in Section 3.1 and is trained multiple times with each dataset.

\begin{figure}
    \centering
    \includegraphics[width=0.85\linewidth]{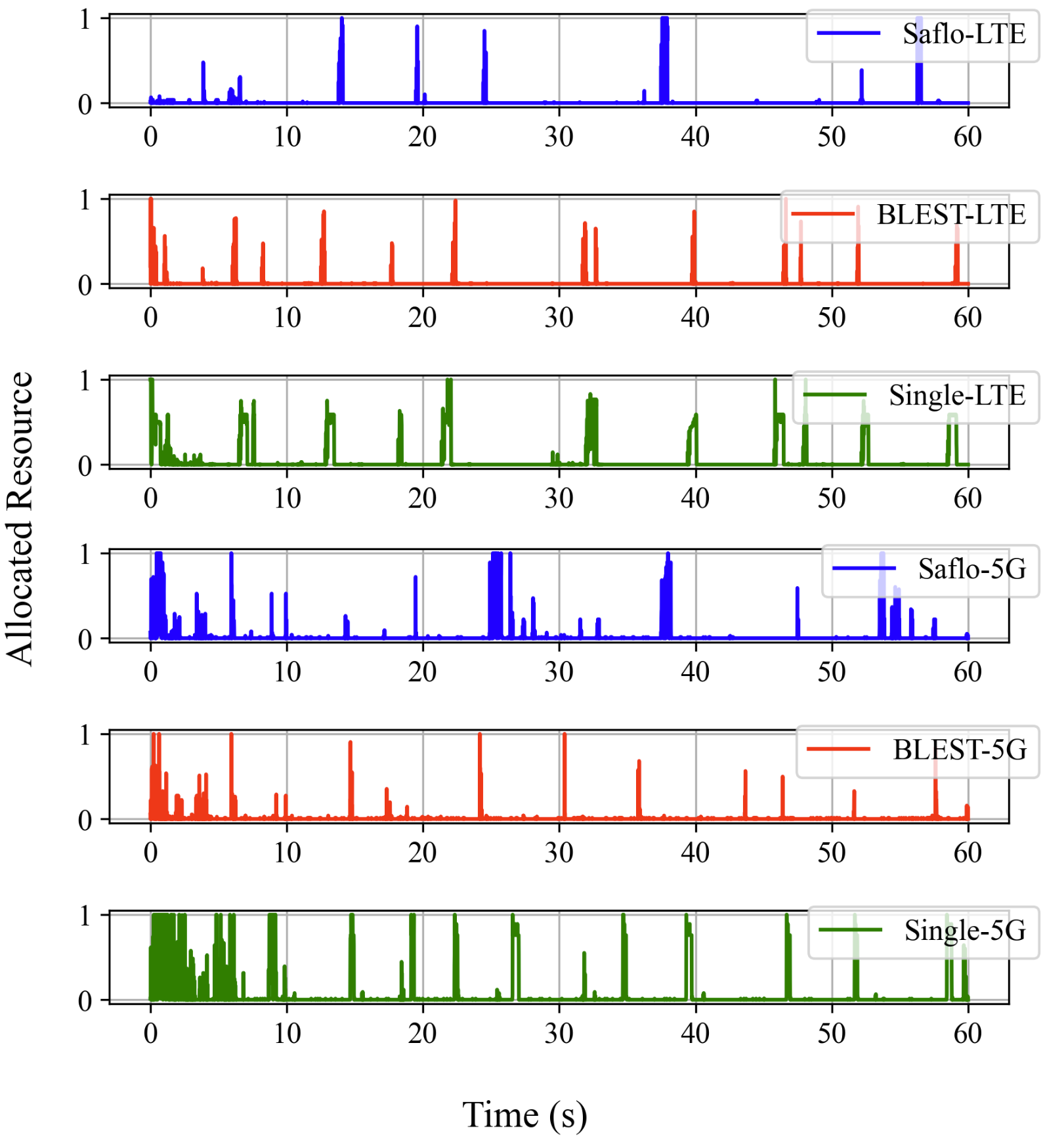}
    \caption{Normalized allocated resources in cellular network over time during YouTube streaming}
\end{figure}

Figure 6 illustrates an example of allocated resources in cellular networks during YouTube video streaming. It represents the traffic pattern captured by the attacker, and the values are normalized to range between 0 and 1 for comparison. When the Saflo scheduler is used, parts of the on-off pattern are absent in the trace because the subflow on cellular networks is randomly disabled. From the attacker's perspective, the traffic pattern generated by the Saflo scheduler includes data occlusion, making the training of the CNN classifier significantly less effective.

The BLEST scheduler reduces traffic sent over cellular networks but still reveals the on-off periods more than the Saflo scheduler. When comparing the allocated resources of the BLEST scheduler in LTE and 5G networks, less traffic is sent over the 5G network. This is due to differences between the eNB and gNB. In our testbed, the LTE eNB is a commercial product, whereas the 5G gNB is an open-source implementation using an SDR, which is less timely and reliable in communication. Consequently, the BLEST scheduler naturally favors using the WiFi subflow more in the 5G network. Meanwhile, when a single LTE or 5G network is used, the traffic pattern clearly exhibits the on-off periods.

Table 1 shows the attack accuracy of video identification across various observation times. Overall, the attack accuracy increases as the observing time increases. When the Saflo scheduler is used, the attack accuracy significantly reduces in all the cases because its traffic pattern has occlusion. The attack accuracy of the Saflo scheduler is higher in 5G networks compared to LTE networks. This can be attributed to two factors. First, the traffic trace in 5G networks is derived from perfect DCI information obtained directly from the gNB log. Second, when only the subflow on the 5G network was enabled, the traffic exhibited a bursty pattern, unlike that observed with the eNB. We believe this behavior stems from the differences between the SDR-based gNB and the commercial eNB.

The BLEST scheduler exhibits at least 50\% higher attack accuracy compared to the Saflo scheduler but lower than that of single LTE/5G networks. The BLEST scheduler reduced the attack accuracy by distributing traffic across the WiFi subflow but was unable to perfectly hide the on-off periods. Additionally, the attack accuracy of the BLEST scheduler is higher in the LTE network than in the 5G network. This is because less traffic is sent over the cellular subflow in the 5G network due to the limitations of the SDR-based gNB. In the singlepath LTE/5G networks, the attack accuracy is over 69\% even if the observing time is 30 seconds. 

\begin{figure*}
    \centering
    \includegraphics[width=1\linewidth]{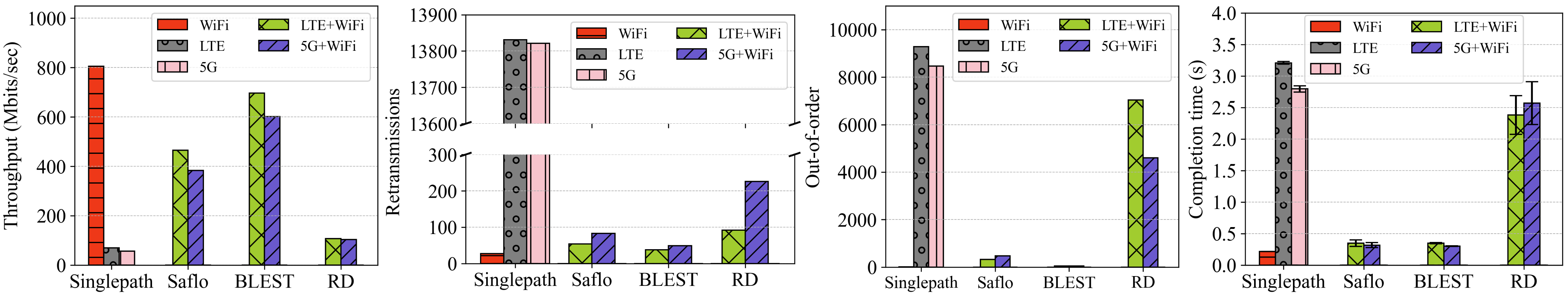}
    \caption{Evaluation of network performance}
\end{figure*}

\begin{table}
\small
\centering
\caption{Attack accuracy of video identification across different observing time}
\begin{tabularx}{\columnwidth}{c*{6}{>{\centering\arraybackslash}X}}
        \hline
        & & 30s & 60s & 90s & 120s \\ 
        \hline
        \multirow{3}{*}{LTE}& Saflo & 0.063 & 0.067 & 0.095& 0.124\\
                            & BLEST & 0.315 & 0.598 & 0.729 & 0.733\\
                            & Singlepath & 0.877  & 0.952& 0.965 & 0.969\\ 
                            \hline
        \multirow{3}{*}{5G} & Saflo & 0.151 & 0.195 &  0.244 & 0.305 \\
                            & BLEST & 0.114 & 0.211 & 0.499 & 0.649 \\
                            & Singlepath & 0.692 & 0.916 & 0.966 & 0.984 \\   
        \hline                      
\end{tabularx}
\end{table}

\begin{table}
\small
\centering
\caption{Top-\textit{k} accuracy of video identification by different schedulers in LTE with 120s observing time}
\begin{tabularx}{\columnwidth}{c*{4}{>{\centering\arraybackslash}X}}
        \hline
                   & Top-1 & Top-2  & Top-3 \\ 
        \hline
        Saflo      & 0.124 & 0.218 & 0.292\\
        RD         & 0.512 & 0.660 & 0.723\\ 
        BLEST      & 0.733 & 0.863 & 0.902\\
        \hline                      
\end{tabularx}
\end{table}

\textbf{Comparison with SOTA in Tor:} We mirrored the operation of TrafficSliver \cite{TrafficSliver} by implementing the RD scheduler with eBPF. TrafficSliver’s approach is a batched random distribution in which it groups traffic (e.g., cells in the context of Tor) and assigns each group randomly to a path. In the RD scheduler, a subflow is chosen randomly each time the scheduler is invoked. Since the Linux kernel's MPTCP scheduler assigns traffic in chunks (i.e., \textit{burst} value), the RD scheduler closely replicates the behavior of TrafficSliver.

We collected the DCI information under the LTE+WiFi environment with the RD scheduler, playing the top 50 YouTube videos 20 times each. In other words, we generated another dataset with the RD scheduler (i.e., \textbf{RD-LTE} dataset). 

Table 2 presents the Top-\textit{k} accuracy of the video identification attack in an LTE+WiFi environment using different MPTCP schedulers. As \textit{k} increases, attack accuracy improves. The Saflo scheduler consistently achieves the lowest attack accuracy. The RD scheduler performs worse than the Saflo scheduler because it tends to reveal on-off periods, with a 50\% chance of assigning traffic to the LTE subflow every on-period. BLEST shows the highest attack accuracy because it lacks randomness in scheduling, making its behavior more predictable over long observation periods.

\begin{table}
    \small
    \caption{Performance of attack detector}
    \centering
    \begin{tabularx}{\columnwidth}{c*{3}{>{\centering\arraybackslash}X}}
        \hline
        Detection delay & Primary classifier  & Secondary classifier \\ \hline
        16.84s & 0.988 & 0.887 \\  
        \hline
    \end{tabularx}
\end{table}
\subsection{Mitigation of User Identification}
\textbf{Performance of attack detector:} The attack detector is essential for mitigating user identification attacks. We evaluated the attack detector's performance in terms of detection delay and classifier accuracy, as shown in Table 3. Detection delay is the time from the initiation of an attack to the moment the detector identifies malicious traffic. The attack was launched 10 times in LTE and another 10 times in 5G, and the attack detector achieved an average detection delay of 16.84 seconds.

We trained the classifiers on 35 minutes of \textit{burst} traces captured in LTE (5 minutes for each of the seven traffic scenarios in Section 3.2) and tested on two separate sets of traces: 35 minutes from LTE and 35 minutes from 5G (a total of 70 minutes). Since all traces were captured by the MPTCP proxy before being transmitted to the user, they share the same patterns regardless of the network. As a result, the primary classifier achieved 98.8\% accuracy, while the secondary classifier achieved 88.7\%.

\textbf{Comparison with other MPTCP schedulers:} To evaluate the mitigation of user identification attack across the different MPTCP schedulers, we assumed the attacker knows what MPTCP scheduler is used for downlink traffic to the user. This is a stronger attacker than that considering the traffic is sent sorely on the cellular network. We collected the DCI information with each MPTCP scheduler under seven traffic scenarios as in Section 3.2. For each traffic scenario, 10 minutes of DCI information was collected (70 minutes for each MPTCP scheduler). Then, we used 80\% as training the attacker's classifier and 20\% for evaluation.   

As shown in Table 4, the Saflo scheduler after attack detection reduced the attack accuracy to below 64\%. Given that this is a balanced binary classification problem, such a low accuracy suggests that the attacker struggles to identify the user's presence. For instance, in the 5G network, the attacker misclassified 8,731 out of 14,725 non-target samples (class 0) as the target (class 1), resulting in a high false positive rate. This is primarily due to the Saflo scheduler's ability to detect malicious traffic and route it to a safer subflow (i.e., WiFi). Before detection, the Saflo scheduler achieved over 90\% attack accuracy, demonstrating that random distribution alone is insufficient for mitigating user identification attacks. Meanwhile, other MPTCP schedulers also showed attack accuracy above 90\%, indicating that simply distributing traffic across subflows is not enough to obscure malicious traffic patterns from the attacker.



\begin{table}[t]
    \small
    \centering
    \caption{Attack accuracy of user identification attacks}
    \begin{tabularx}{\columnwidth}{c*{6}{>{\centering\arraybackslash}X}}
        \hline
                               & LTE & 5G \\
                               \hline
          Saflo after detection  & 0.595 & 0.632 \\
       Saflo before detection & 0.923 & 0.948 \\
                         BLEST & 0.976 & 0.922 \\
                            RD & 0.976 & 0.982 \\
                    Singlepath & 0.985 & 0.998 \\
                               \hline
    \end{tabularx}
\end{table}

        

\subsection{Network Performance}

We evaluated network performance using four metrics: throughput, retransmissions, out-of-order TCP packets, and completion time. For throughput, retransmissions, and out-of-order TCP packets, we utilized \textit{iperf3} to generate saturated traffic over a single MPTCP socket from the proxy to the user for 10 minutes. To measure completion time, we deployed an HTTP server and loaded a 15.6 MB web page. The web page was accessed 200 times for each instance to calculate the average completion time.

Figure 8 shows the network performance of singlepath networks and MPTCP schedulers. In the figures, the singlepath with WiFi showed the best performance, while the singlepath with the cellular networks showed the worst; notably, there is a significant difference in the capabilities of WiFi 6 and private cellular networks. This disparity influenced the performance of MPTCP schedulers. Specifically, when the WiFi subflow was used more frequently than the cellular subflow, better performance was observed intuitively. 

Overall, the Saflo scheduler outperformed the RD scheduler due to its ability to account for linger time, effectively reducing HoL blocking. The Saflo scheduler achieved a significant reduction in out-of-order packets compared to the RD scheduler—89\% in 5G networks and 94\% in LTE networks. This highlights the Saflo scheduler's superior ability to mitigate HoL blocking.

However, the Saflo scheduler could not surpass the performance of the BLEST scheduler because of additional security-related operations. In particular, there were moments when only the subflow on the cellular network was active, causing bottlenecks that limited the Saflo scheduler's performance.

\section{Discussion}
\textbf{Why is it not a good idea to rely on WiFi only?}
Although our focus was on defending against attack models in cellular networks, traffic analysis attacks can exist in WiFi networks. The vantage point for such attacks can extend from the WiFi access point to any routers or switches along the communication route \cite{WiFi-web,WiFi-profile, EncryptedHTTP, Beauty}. The Saflo scheduler can mitigate video identification attacks and detect user identification attacks in WiFi networks as well. Furthermore, with optimized parameters—such as the subflow manager's operation interval, $P_{max}$, and $P_{min}$—it can also likely mitigate other traffic analysis attacks. An important future work would be demonstrating the Saflo scheduler's ability to mitigate traffic analysis attacks across various networks and attack types.

\textbf{What if the attacker can simultaneously monitor both cellular and WiFi networks?} From an attacker's perspective, it is challenging to link the UE's RNTI on the cellular network with the IP address on the WiFi network. Even if the attacker uses a DCI sniffer for the cellular network and gains a vantage point on the WiFi network, they are unable to effectively correlate traffic metadata between the two networks. While there may be potential methods to establish such a link, we have not identified a viable approach, particularly when the WiFi path is secured by an IPSec tunnel implemented through the N3IWF function.

\textbf{Limitation of the current open-source sniffers:} As discussed in Section 3, we were unable to deploy the open-source sniffers against the MNO's eNB and gNB in our area. For LTE networks, the primary eNBs in our region operate in MIMO mode. While LTESniffer supports MIMO configurations up to 2×2, the eNBs in our area employed MIMO configurations beyond 2×2. Although we identified an eNB on a different band that did not use MIMO, it operated in TDD mode. While the LTESniffer can capture DCIs from TDD eNB signals, it cannot decode the PDSCH in TDD mode. This limitation becomes critical due to the use of CA. As mentioned in Section 2, overcoming CA requires the sniffer to decode PDSCH data and infer the traffic transmitted on SCells. Unfortunately, the LTESniffer is incapable of performing this operation with TDD eNBs.

The 5GSniffer is unable to capture DCI information from gNBs operating in MIMO mode. Since the primary gNBs in our area use MIMO configurations, we were unable to sniff DCIs from them. While we identified gNBs that do not use MIMO, the 5GSniffer failed to capture the complete PDCCH from these gNBs due to its fixed master clock. In other words, part of the PDCCH was transmitted on a frequency that the 5GSniffer could not observe when it was centered on the SS Block (SSB) frequency. This occurred due to the flexible SSB and PDCCH configuration in 5G networks.

Although we found that it is possible to deploy the user identification attack against a specific gNB in our area by forcing the phone to connect to it, this approach required discarding portions of the collected data when the decoded number of DCIs is small, as mentioned in Section 3.2. This made it difficult to use the 5GSniffer for evaluation because our countermeasure naturally results in fewer DCIs. Thus, we opted to evaluate our countermeasure in our testbed to ensure clear and reliable experimental results.

\textbf{Limitation of the Saflo scheduler:}
The main limitation of the Saflo scheduler is its exclusive support for TCP traffic, as it depends on MPTCP. This means that UDP and other non-TCP traffic are routed only through the default network interface. While most traffic analysis attacks target TCP traffic, such as video streaming, messaging, and web browsing, variations of these attacks could exploit UDP traffic. Extending the Saflo scheduler to handle UDP traffic via a QUIC proxy \cite{QUIC-proxy} is an important direction for future research.



\section{Conclusions}
We present the Saflo scheduler, an eBPF-based MPTCP scheduler designed to defend against traffic analysis attacks in cellular networks. Specifically, it protects against video identification attacks \cite{VideoID} and user identification attacks \cite{5GSniffer}, which have shown its feasibility in real-world LTE/5G networks. The Saflo scheduler utilizes eBPF to operate across both kernel and user spaces, allowing security-related computations and machine learning-based attack detection to be offloaded to user-space programs. The Saflo scheduler also demonstrates how complex tasks, like machine learning models, can be integrated into kernel-level operations using eBPF to enhance network security. Our evaluation shows that the Saflo scheduler dramatically lowers video identification attack accuracy from 96.9\% to 12.4\% and user identification attack accuracy from 99.4\% to 59.5\%. Moreover, it surpasses single-path cellular networks and the SOTA approach in the Tor network \cite{TrafficSliver} in terms of throughput, out-of-order packets, retransmission, and completion time.

\bibliographystyle{ACM-Reference-Format}
\bibliography{Main}

\appendix

\section{Operation of eBPF scheduler}
Algorithm 2 outlines how the eBPF scheduler selects subflows for data transmission. The process begins by initializing the \textbf{sendInfo} structure, which stores information about the selected subflow. The scheduler then iterates over all subflows belonging to a single MPTCP connection. It is important to note that the kernel invokes the eBPF scheduler whenever data transmission needs to be scheduled for a specific MPTCP connection. As a result, the subflow information for that connection is provided as a set of \textbf{mptcp\_subflow\_context} structures. For simplicity, we refer to this structure as \textbf{subflow\_context} throughout Algorithm 2 and this paper.

During each iteration, the scheduler attempts to retrieve the entity from the eBPF map corresponding to the subflow's unique key, which is composed of the subflow's token, remote ID, and local ID. If the entity cannot be retrieved, the scheduler generates a new \textbf{map\_entity} with appropriate values and inserts it into the eBPF map using the subflow's key. This ensures that every subflow has a corresponding entry in the eBPF map.

Once the map entity is retrieved or generated, the eBPF scheduler calculates the current \textit{linger\_time} for the subflow. If the subflow is enabled, as indicated by setting \textbf{map\_entity}.\textit{enabled} to True, the scheduler compares the subflow's \textit{linger\_time} with the \textit{linger\_time} stored in \textbf{sendInfo}. If the subflow's \textit{linger\_time} is shorter, the subflow's \textbf{subflow\_context} structure and \textit{linger\_time} replace the existing information in \textbf{sendInfo}.

\begin{algorithm}[t]
\caption{Subflow selection in eBPF scheduler}
\definecolor{lightblue}{RGB}{173, 216, 230} 
\sethlcolor{lightblue}
\begin{algorithmic}[1]
\REQUIRE subflows[\textit{i}]: an array of \textbf{subflow\_context}
\STATE /* Initialize \textbf{sendInfo} */
\STATE \textbf{sendInfo}.$selectedSF \gets NULL$ \\ 
\STATE \textbf{sendInfo}.$linger\_time \gets -1$
\STATE /* Iterate over subflows */
\FOR{each \textbf{subflow\_context} in subflows}
    \STATE \textbf{map\_entity} = Read map entity of \textbf{subflow\_context}
    \STATE /* Insert new entity to the map if it does not exist */
    \IF{\textbf{map\_entity} is empty}
        \STATE Create new \textbf{map\_entity} for \textbf{subflow\_context}
        \STATE Insert \textbf{map\_entity} to the eBPF map.
    \ENDIF
    \STATE $linger\_time \gets \frac{\textbf{subflow\_context}.wmem}{\textbf{subflow\_context}.pace}$
    \STATE /* Subflow selection is performed below */
    \IF{linger\_time $<$ \textit{\textbf{sendInfo}}.linger\_time \&\& \\ \textbf{map\_entity}.\textit{enabled}==True}
        \STATE $\textbf{sendInfo}.selectedSF \gets \textbf{subflow\_context}$
        \STATE $\textbf{sendInfo}.linger\_time \gets linger\_time$
    \ENDIF
    \STATE /* Update the map entity with new information */
    \STATE Update \textbf{map\_entity} with \textit{linger\_time}, \\ \textbf{subflow\_context}.\textit{wmem}, \textbf{subflow\_context}.\textit{pace}
    \STATE Replace the existing entity in the eBPF map with \textbf{map\_entity}
\ENDFOR
\RETURN \textbf{sendInfo}
\end{algorithmic}
\end{algorithm}

At the final step of each iteration, the eBPF scheduler updates the \textbf{map\_entity} with the newly obtained information, including \textit{linger\_time}, \textit{wmem}, and \textit{pace}. It then replaces the existing entry for the subflow in the eBPF map with the updated \textbf{map\_entity}.

The subflow selection described in Algorithm 2 does not apply to retransmissions. For retransmissions, we have retained the existing BLEST scheduling function implemented in the kernel. This decision was based on the fact that the existing function prioritizes the fastest possible transmission, which we consider a more appropriate approach for retransmissions.

\section{CNN Classifier for Attack Detection}
The primary and secondary classifiers for attack detection have the same structure which is 1D CNN models designed for sequence-based binary classification. It begins with an input layer that accepts data sequences of 1000 long (i.e., 10 seconds) and normalizes the input using a Lambda layer. Two convolutional layers follow, each using a filter size of 150 and a kernel size of 40. These layers are combined with batch normalization and LeakyReLU activations. Dropout (rate of 0.4) and average pooling layers are applied after each convolution to reduce overfitting and downsample the feature maps while preserving essential information.

After the convolutional feature extraction, the network flattens the data and passes it through several fully connected layers. The dense layers reduce feature dimensions from 512 to 128, each followed by a dropout layer to improve generalization. Finally, a sigmoid-activated output layer predicts the binary classification result.


\end{document}